\documentclass[conference]{IEEEtran}
\usepackage{cite}
\usepackage{amsmath,amssymb,amsfonts}
\usepackage{graphicx}
\usepackage{textcomp}
\usepackage{xcolor}
\usepackage{cuted}
\usepackage{booktabs}
\usepackage{mathtools}
\usepackage{bm}
\usepackage{array}
\usepackage{float}
\usepackage{algorithm}
\usepackage[noend]{algpseudocode}
\usepackage{multirow}
\usepackage[export]{adjustbox}
\usepackage{titlesec}

\usepackage[font=footnotesize,skip=2pt]{caption} % 'skip' is vertical space between figure and caption
\usepackage{eso-pic}

\usepackage{makecell}
\usepackage{threeparttable}
\usepackage{comment}

\usepackage{hyperref}
\hypersetup{
    colorlinks=true,
    linkcolor=blue,
    filecolor=magenta,
    urlcolor=blue,
    citecolor=blue,
}
\algdef{SE}[DOWHILE]{Do}{doWhile}{\algorithmicdo}[1]{\algorithmicwhile\ #1}%
\algdef{SE}[REPEAT]{Repeat}{Until}{\algorithmicrepeat}[1]{\algorithmicuntil\ #1}%
\def\BibTeX{{\rm B\kern-.05em{\sc i\kern-.025em b}\kern-.08em
    T\kern-.1667em\lower.7ex\hbox{E}\kern-.125emX}}

\AddToShipoutPictureBG{%
    \AtPageUpperLeft{%
        \raisebox{-1cm}{% adjust vertical position
            \makebox[\paperwidth]{\hfill
                \parbox{0.9\paperwidth}{\centering
                    \textbf{3rd International Conference on Computational Intelligence and Network Systems (CINS 2025),}\\
                    \textbf{BITS Pilani, Dubai Campus, Dubai, UAE}%
                }
            \hfill}%
        }%
    }%
}

\begin{document}
% paper ID : 170
\title{A Quantum-Secure and Blockchain-Integrated E-Voting Framework with Identity Validation}
% \title{QVote: Quantum Secure E-Voting Framework with Blockchain and Identity Validation}
% The footnote here is for the starred affiliation and shared email block
\thanks{Corresponding author email: \textit{subramaniyaswamy.v@vit.ac.in}}
% Using the standard IEEEtran six-column author block layout for the six authors
\author{
    \IEEEauthorblockN{Ashwin Poudel}
    \IEEEauthorblockA{\textit{School of CS and Engineering,}\\
        \textit{Vellore Institute of Technology, India}\\
        \textit{ashwin.poudel2020@vitalum.ac.in}}
    \and
    \IEEEauthorblockN{ Utsav Poudel}
    \IEEEauthorblockA{\textit{School of CS and Engineering,}\\
        \textit{Vellore Institute of Technology, India}\\
        \textit{utsav.poudel2021@vitstudent.ac.in}}
    \and
    \IEEEauthorblockN{Dikshyanta Aryal}
    \IEEEauthorblockA{\textit{Institute of Engineering,}\\
        \textit{Tribhuvan University, Pokhara, Nepal}\\
        \textit{pas076bei010@wrc.edu.np}}
    \and
    \IEEEauthorblockN{Anuj Nepal}
    \IEEEauthorblockA{\textit{Universal Higher Education, Melbourne}\\
        \textit{Deakin University, Geelong, Australia}\\
        \textit{anepal@deakin.edu.au}}
    \and
    \IEEEauthorblockN{Pranish Pathak}
    \IEEEauthorblockA{\textit{Academy of Interactive Technology,}\\
        \textit{Sydney, Australia}\\
        \textit{12275@ait.nsw.edu.au}}
    \and
    \IEEEauthorblockN{Subramaniyaswamy V}
    \IEEEauthorblockA{\textit{School of CS and Engineering,}\\
        \textit{Vellore Institute of Technology, India}\\
        % The email is moved to the \thanks footnote as the * indicates the corresponding author
        \textit{subramaniyaswamy.v@vit.ac.in}}
}
\maketitle
% --- Add IEEE copyright notice here ---
\begin{center}
\footnotesize
% During review:
This work has been submitted to the IEEE for possible publication.\\
% After acceptance: © 2025 IEEE. Personal use permitted. Permission from IEEE required...
\end{center}

\begin{abstract}
The rapid growth of quantum computing poses a threat to the cryptographic foundations of digital systems, requiring the development of secure and scalable electronic voting (e-voting) frameworks. We introduce a post-quantum secure e-voting architecture that integrates Falcon lattice-based digital signatures, biometric authentication via MobileNetV3 and AdaFace, and a permissioned blockchain for tamper-proof vote storage. Voter registration involves capturing facial embeddings, which are digitally signed using Falcon and stored on-chain to ensure integrity and non-repudiation. During voting, real-time biometric verification is performed using anti-spoofing techniques and cosine similarity matching. The system demonstrates low latency and robust spoof detection monitored through Prometheus and Grafana for real-time auditing. The average classification error rates (ACER) are below 3.5\% on CelebA-Spoof and under 8.2\% on the Wild Face Anti-Spoofing (WFAS) dataset. Blockchain anchoring incurs minimal gas overhead, approximately 3.3\% for registration and 0.15\% for voting,  supporting system efficiency, auditability, and transparency. The experimental results confirm the system's scalability, efficiency, and resilience under concurrent loads. This approach offers a unified solution to address key challenges in voter authentication, data integrity \& quantum-resilient security for digital systems.
\end{abstract}


\section{Introduction}
The evolution of digital governance and civic participation has amplified the need for secure and transparent electronic voting (e-voting) systems.
%The confidentiality and reliability of electoral processes are very important as society transitions to digital platforms.
However, the increase in cyber threats, such as data breaches, deepfakes, and algorithmic manipulations, poses substantial risks to the credibility of democratic systems \cite{de}. Conventional cryptographic techniques, such as Rivest-Shamir-Adleman (RSA) and Elliptic Curve Cryptography (ECC), face challenges from emerging quantum computing advancements that undermine the widely used encryption standards \cite{Sahoo}.
%This evolving threat landscape mandates the development of next-generation e-voting systems that integrate quantum-resistant security with usability, scalability, and robust voter authentication.

Most existing e-voting systems depend on centralized architectures and traditional cryptographic techniques, which are vulnerable to single points of failure, vote tampering, and potential future quantum attacks. Furthermore, these systems lack robust user authentication, increasing the risk of voter impersonation, unauthorized access, and electoral fraud. 

To address these challenges, there is a need for a next-generation e-voting framework that integrates quantum-resistant encryption, blockchain-based decentralized storage, and biometric face recognition. This guarantees that votes are cast securely, stored verifiably, and authentically attributed. Our main contributions are: 1. A tamper-proof e-voting framework using post-quantum cryptography (PQC) to safeguard voting data. 2. A lightweight anti-spoofing facial recognition model based on MobileFaceNet with Adaface, to provide efficient and secure real-time voter authentication on resource-constrained devices.
3. A unified architecture combining Fast-Fourier Lattice-based Compact Signatures over NTRU (Falcon) for quantum encryption, blockchain for transparent vote storage, and MobileFaceNet for biometric verification.

The remainder of this paper is organized as follows: Section II reviews related work. Section III outlines the system architecture.
%, detailing the integration of Falcon post-quantum encryption, MobileFaceNet-based facial recognition, and blockchain-based vote storage. 
Section IV describes the proposed methodology.
%, including cryptographic key generation, voter registration, anti-spoofing mechanisms, and smart contract implementation. 
Section V presents the experimental results. Finally, Section VI concludes with future directions.
%for enhancing the scalability, privacy, and ethical deployment of biometric e-voting systems.
\section{Related Work} \label{sec:RW}
 Traditional e-voting systems relied on centralized servers and limited cryptographic techniques, making them vulnerable to data breaches and single points of failure \cite{Choi_2021}. To address these limitations, modern e-voting systems have implemented drastic changes in two-factor authentication, cryptographic hashing, and blockchain-based systems to improve transparency and security \cite{Alvi}. Despite the changes, these systems remain vulnerable to evolving cyber threats, particularly from quantum computing \cite{Mahanayak2023}.
 
 Hybrid blockchain architectures that merge public and private chains have been proposed to enhance security and performance \cite{JAYAKUMARI2024102}, while lightweight blockchain solutions have been explored in domains such as the Internet of Vehicles to reduce overhead \cite{10674095}. Even though these innovations focused on verification, identity authentication, and the system’s reliability, most of the solutions address isolated components rather than providing an integrated approach.
 %that relies on quantum encryption with biometric technology.

\begin{table}[htp]
\centering
\caption{Comparison of existing E-Voting systems with our Proposed Model}
\label{comparision}
\resizebox{\columnwidth}{!}{
\begin{tabular}{l c c c c c c c c c}
\hline
Properties &  \cite{Choi_2021} & \cite{Alvi} & \cite{fi16110388} & \cite{https://doi.org/10.1111/exsy.13694} & \cite{Mahanayak2023} & \cite{10.1109/JIOT.2025.3542996} & \cite{10.1117/12.3054179} & \cite{9425815} & \textbf{Proposed} \\
 & & &  &  &  &  &  &  & \textbf{Framework} \\
\hline\hline
Unforgeability     & $\checkmark$ & $\checkmark$ & $\checkmark$ & $\checkmark$ & $\checkmark$ & $\times$ &$\times$  &$\checkmark$ & $\checkmark$ \\
Anonymity          &  $\checkmark$       & $\checkmark$             & $\checkmark$             & $\checkmark$             &  $\checkmark$            & $\times$  & $\times$ & $\checkmark$ & $\checkmark$ \\
Correctness       & $\checkmark$ & $\checkmark$ & $\checkmark$ & $\checkmark$ & $\checkmark$ & $\checkmark$ &$\checkmark$  &$\checkmark$ & $\checkmark$ \\
Verifiability       & $\checkmark$ & $\checkmark$ & $\checkmark$ & $\times$ & $\checkmark$ & $\times$ &$\times$  &$\checkmark$ & $\checkmark$ \\
Immutability     & $\checkmark$ & $\checkmark$ & $\checkmark$ & $\checkmark$ & $\checkmark$ & $\times$ &$\checkmark$  &$\checkmark$ & $\checkmark$ \\
Robustness        & $\times$ &$\times$ & $\checkmark$ & $\checkmark$ & $\times$ & $\checkmark$ &$\checkmark$  &$\checkmark$ & $\checkmark$ \\
Fault tolerance    & $\times$ &$\times$ & $\checkmark$ & $\checkmark$ & $\times$ & $\times$ &$\times$  &$\checkmark$ & $\checkmark$ \\
Scalability         & $\times$ & $\times$ & $\checkmark$ & $\checkmark$ & $\checkmark$ & $\checkmark$ &$\checkmark$  &$\checkmark$ & $\checkmark$ \\
Quantum Encryption & $\times$ &$\times$ & $\times$ & $\times$ & $\checkmark$ & $\times$ &$\times$  &$\times$& $\checkmark$ \\
\hline
\end{tabular}
}
\end{table}

The advancements in quantum computing, as shown in Table~\ref{comparision}, have rendered traditional cryptographic algorithms, such as RSA and ECC, obsolete due to the emergence of various quantum algorithms. To address this problem, the PQC scheme, Falcon, is promising due to its hard lattice-based Short Integer Solution (SIS) problem \cite{falcon_sign_2025}. It is resistant to both classical and quantum attacks, featuring a compact signature size, faster key generation, and secure digital signature capabilities, making it a prominent candidate for resource-constrained environments, such as e-voting applications \cite{Prajapat2025Quantum}. The study by Liu et al. (2025) \cite{10.1109/JIOT.2025.3542996} presents a new complex system, Sine-Cosine Coupled Mapping Lattice (SCCML), which is used for encrypting multiple facial images. The system employs a digital separation loop to enhance the privacy of encrypted biometric data, but SCCML is still immature for a real-world e-voting system. 
Facial recognition remains popular due to its non-intrusive and remote nature. Despite significant progress in secure e-voting, the current solution only focuses on isolated components rather than offering an integrated framework. 

To build a truly resilient and tamper-proof e-voting system, quantum-resistant encryption must be complemented by robust biometric authentication and anti-spoofing mechanisms. MobileFaceNet is the most promising model in this domain, which is a lightweight and highly accurate convolutional neural network (CNN) tailored for real-time face verification on mobile and embedded devices \cite{10.1117/12.3054179}. MobileFaceNet employs depthwise separable convolutions and residual bottlenecks to improve computational efficiency. It integrates Red-Green-Blue (RGB) input with alternative multi-modal sensing techniques such as Near-Infrared (NIR), depth, and thermal imaging, significantly enhancing spoof detection against advanced three-dimensional (3D) mask attacks. While widely used in biometrics, its role in e-voting remains limited. A blockchain-based, privacy-preserving face authentication system using Trusted Execution Environments (TEEs) and distributed deep learning has been proposed. Similarly, Peelam et al. (2025) \cite{https://doi.org/10.1111/exsy.13694} introduced DemocracyGuard, a blockchain-based voting platform incorporating facial recognition for authentication via Ethereum- smart contracts. However, this lags behind encryption techniques. AdaFace is a recent face recognition framework that enhances traditional margin-based Softmax (Loss function) approaches by introducing a quality-adaptive margin. This margin dynamically adjusts based on the image’s feature norm, which serves as a proxy for image quality \cite{Kim_2022_CVPR}. The model uses ResNet-50 to create a 512-dimensional facial embedding, encrypted with Falcon Quantum cryptography and securely stored. Input images are detected, aligned, and resized to 112×112 pixels \cite{8953658}. Although several projects have implemented blockchain in electoral systems, challenges such as voter privacy, data interoperability, and system scalability continue to hinder real-world adoption \cite{Choi_2021} \cite{JAYAKUMARI2024102}.In a public e-voting system, the possibility of irreversible privacy violations can be increased if biometric data are improperly secured. 

%Table~\ref{comparision} showcases different privacy approaches previously used in e-voting systems. 
%We list their major privacy approaches and challenges they face in the real world. 
To address these gaps, we propose a fully integrated, quantum-secure, and biometric-enabled e-voting framework. By combining Falcon-based quantum encryption, MobileFaceNet for anti-spoofing, Adaface for facial recognition, and a permissioned blockchain for transparent vote recording, it contributes a resilient and scalable solution for secure digital democracy.

\section{Methodology}\label{sec:methodology}
This section outlines the technical workflow for designing and developing the e-voting system. The methodology is structured into five distinct phases: Environment setup, PQC with Falcon, Authentication and Anti-spoofing, Smart contract development, and System integration.

\subsection{ Experiment Setup}
The blockchain infrastructure is initiated through Ethereum, while Hardhat is a core development environment. HardHat provides a built-in local Ethereum network that runs as a JSON-RPC / WebSocket server, enabling rapid compilation, deployment, and testing of smart contracts on a simulated chain \cite{hardhat_2025}. This local node comes with pre-funded test accounts and mines a block for each transaction, facilitating a convenient development and debugging environment. We use ethers.js in conjunction with Hardhat to programmatically compile and deploy contracts, and tend transactions via the JSON-RPC interface \cite{ethersjs_v6_2025}. Ethereum was chosen as the platform due to its mature smart contract ecosystem, robust security model, and extensive developer support.

\begin{figure}[H]
    \centering
    \includegraphics[width=\columnwidth]{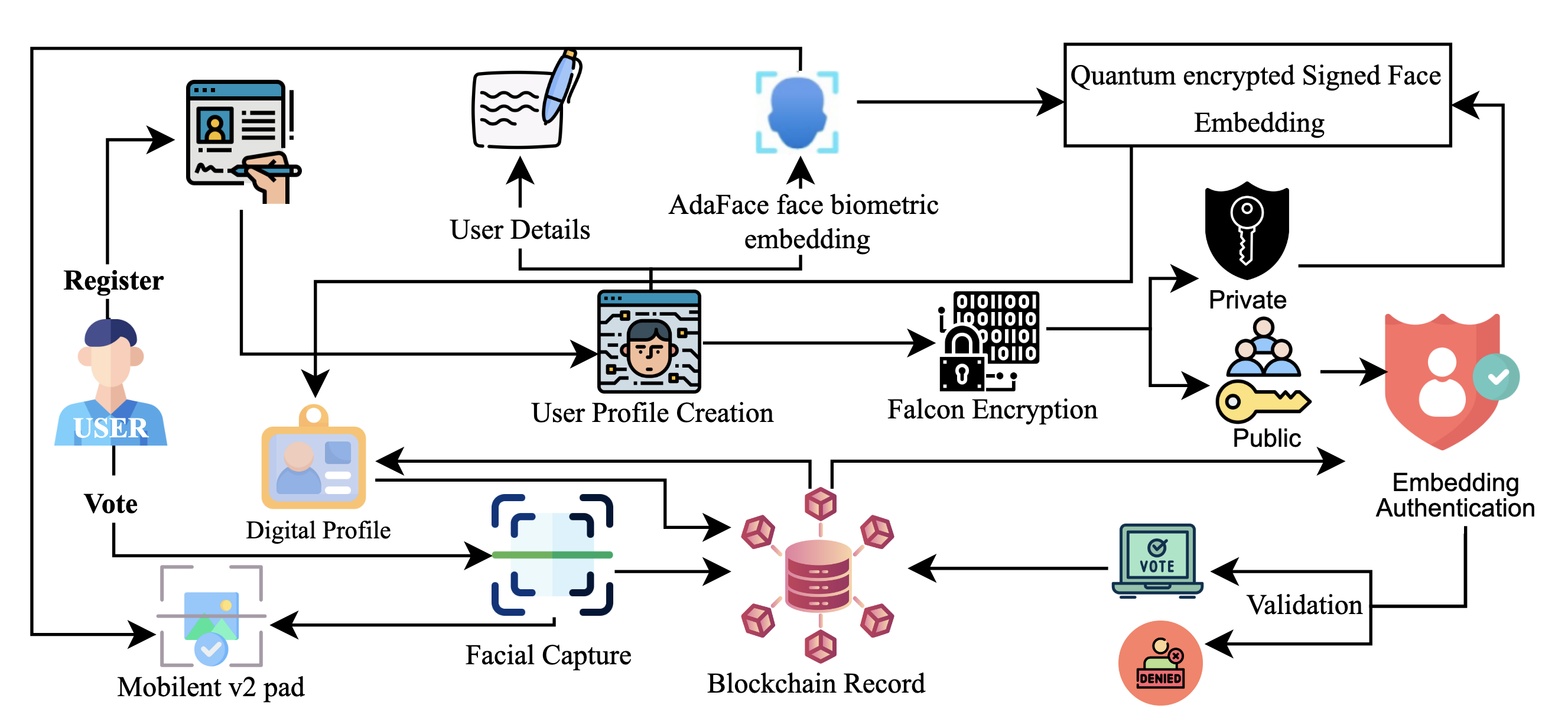} 
    \caption{Architecture using Falcon PQC and MobileNetV3 anti-spoofing}
    
    \label{fig:evoting_system1}
\end{figure}

Figure~\ref{fig:evoting_system1} illustrates a blockchain-based e-voting system using FaceNet for biometric facial recognition and Falcon, a PQC algorithm for encryption. During registration following info is taken: Full name, Phone number, Date of Birth, Citizenship Number, Address, and Face Embeddings.
To ensure the authenticity, we utilize a Falcon-Crypto module. The private key is used to digitally sign the generated face embedding, resulting in a signed face embedding that securely binds the citizen’s identity to their digital profile. The signed embedding and registration details are also stored in the blockchain, providing a decentralized and tamper-proof ledger that ensures the integrity and immutability of the registered data.

During voting, the face will be scanned again. The generated embedding is compared with the previously signed embedding stored on the blockchain. The system verifies the authenticity of the signed face embedding using the Falcon public key right before the comparison with other voter details. If the embedding is matched, it confirms the voter's identity and grants access to cast the vote; otherwise, the voting process will be terminated immediately. After voting concludes, the casted vote is counted and stored again to prevent multiple voting attempts by a single citizen. 

\subsection{Post-Quantum Cryptography with Falcon}
 % \begin{figure}[h]
 %     \centering
     
 %     \includegraphics[angle=270, trim=6cm 0cm 6cm 0cm, keepaspectratio, clip, width=1\columnwidth]{quantumPQC.pdf} 
 %     \caption{Quantum PQC Architecture}
 %     \label{fig:quantum_architecture}
 % \end{figure}

% Figure~\ref{fig:quantum_architecture} 

%demonstrates that 
A lattice-based cryptographic algorithm, Falcon, is used. The generated private key is stored in encrypted files or Hardware Security Modules (HSMs) to prevent unauthorized access, while public keys are distributed among the nodes to ensure integrity.
The private key generates a signature that securely encodes the data, whereas the public key is used to verify the encoded data.

\begin{comment}
\begin{algorithm}
\caption{Lattice-Based Key Generation and Voter Registration}
\label{algo:falcon_voter_registration}
\begin{algorithmic}[1]
\State \textbf{Generate Falcon Key Pairs}

\Function{KEYGEN}{$N, p, q, f, g$}
    \State $fq \gets f^{-1} \bmod q$
    \Ensure $f \cdot fp \bmod p = 1$ and $f \cdot fq \bmod q = 1$
    \State $Pv \gets (f, fq)$
    \State $h \gets p \cdot fq \cdot f \bmod q$
    \State $Pb \gets h$
    \State \Return $(Pv, Pb)$
\EndFunction

\Function{REGISTER}{$VoterAddress, Fe, Pv$}
    \State $H \gets \Call{Hash}{Fe}$
    \State $(s_1, s_2) \gets \Call{falcon.sign}{H, Pv}$
    \State $signed\_embedding \gets (s_1, s_2)$
    \State $Block \gets \Call{new}{BlockStruct}$
    \State $Block.Voter \gets VoterAddress$
    \State $Block.SignedEmbedding \gets signed\_embedding$
    \State $Block.TimeStamp \gets \Call{getTime}{}$
    \State $BlockDigest \gets \Call{generateBlockSignature}{Block}$
    \State $Block.Signature \gets \Call{falcon.sign}{BlockDigest, Pv}$
    \State \Call{AppendToLedger}{Block}
    \State $voter.isRegistered \gets \text{true}$
    \State \Return Success
\EndFunction

\end{algorithmic}
\end{algorithm}
    
\end{comment}

%Algorithm~\ref{algo:falcon_voter_registration} illustrates that the lattice-based cryptographic algorithm called Falcon (KEYGEN(N,p,q,f,g)) generates a key pair, i.e., private and public keys. The private key is used to generate a signature that securely encodes the data. Similarly, the public key is used to verify the encoded signature, thus ensuring the integrity of the data.

Falcon is used for a combined signature approach, where the private key signs and encrypts face embeddings, and the public key enables secure extraction for comparison. During voter registration, embeddings are hashed and signed, with a 'blockstruct' created to store the address, signed embedding, and timestamp. The block digest is also signed using Falcon with the same private key.
\vspace{-1em}
\begin{equation}
\text{Sgn}(sk, m) = (\mathbf{r}, \mathbf{s})
\end{equation}
\begin{equation}
\text{such that} \quad \mathbf{s} = f^{-1}(H(\mathbf{r}, \mathbf{m})) \quad \text{and} \quad \|\mathbf{s}\|_2 \leq \beta
\end{equation}

Equation 1 begins by sampling a random value called salt (r), ensuring that the same message produces different signatures on each execution \cite{cryptoeprint:2024/1769}. The algorithm then enters a loop that computes a hash of the salt and attempts to find a value that, when passed through a function f, yields the hash value. This is done using the inverse function $f^{-1}$, and the ‘s’ is chosen randomly from all possible hash preimages. The loop continues till the Euclidean length of ‘s’ is less than or equal to the threshold $\beta$, which ensures that ‘s’ is small enough to prevent information leakage about the secret key. Once all the conditions are met, the final signature $\sigma$ is the output.

\subsection{Biometric Authentication and Anti-Spoofing}

\begin{figure}[h]
    \centering
    \includegraphics[width=1\columnwidth]{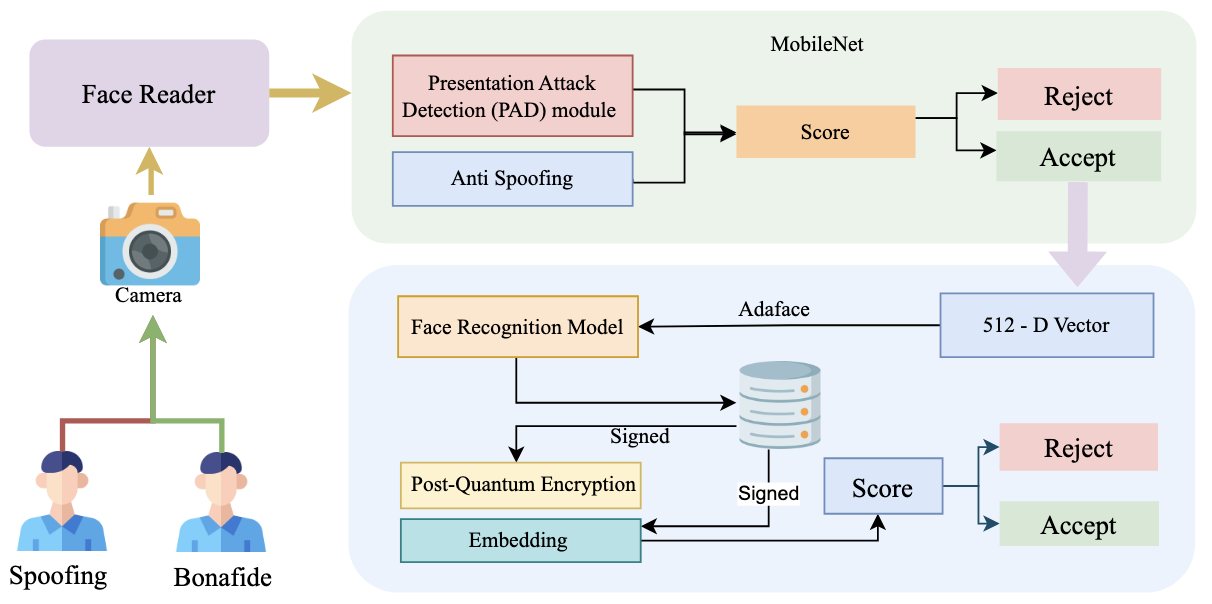} 
    \caption{Anti-spoofing  model with MobileNetV3 and AdaFace for  verification.}
    \label{fig:face_recognitions}
\end{figure}

The proposed biometric authentication system integrates AdaFace-based face recognition with a lightweight anti-spoofing pipeline constructed on MobileNetV3. As shown in Figure~\ref{fig:face_recognitions}, both subsystems work together, first validating liveness using MobilenetV3 to prevent presentation attacks, then verifying user identity through face recognition.

The face authentication is 
%~\ref{algo:face_authentication}
 initiated by preprocessing the input facial image through normalization, alignment, and resizing to a canonical 112×112 format, ensuring geometric and photometric consistency across samples. This standardized input is passed through a presentation attack detection (PAD) model based on MobileNetV3, augmented with Convolutional Block Attention Modules (CBAM) and Channel-Deep Convolutions (CDC), which enhances its sensitivity to spoofing cues such as texture inconsistencies, motion artifacts, and spectral distortions. Upon receiving the PAD output, the system immediately rejects the sample if classified as a spoof (e.g., print, replay, or deepfake). For inputs labeled “live”, the image proceeds to the AdaFace module. The model generates a 512-dimensional face embedding with a quality-adaptive margin for robustness and verifies identity by comparing cosine similarity against a database threshold.

The embeddings produced by the AdaFace module are digitally signed using Falcon. These signed embeddings are stored on-chain in a permissioned blockchain environment. The integration of PQC and biometric authentication within a blockchain ledger provides a secure and auditable framework, ensuring that voter biometrics remain encrypted throughout computation.

Similarly, the signature verification mechanism is achieved when the retrieved signed embedding is validated against the input embedding using a public key. A digest H was generated based on the provided face embedding. The signature components s1 and s2 are computed using an equation (s1 + s2) · Pb mod q, which is further compared with hash H to ensure that the signature matches the original embedding.

% \begin{figure}[H]
%     \centering
%     \includegraphics[trim=0cm 0.5cm 0cm 0.5cm, clip, width=\columnwidth]{lossCalculation.pdf}
%     \caption{Anti-spoofing pipeline built on MobileNetV3, operating alongside AdaFace for secure face verification}
%     \label{fig:face_recognition}
% \end{figure}

\subsection{Smart Contract Development}
Smart contracts, written in Solidity, utilize mappings and structs to associate Ethereum addresses with encrypted face embeddings and track whether each user has casted a vote or not. Events like “voterRegistered" and “castVote" are emitted for transparency and logging \cite{solidity_docs_2016}. We compile the smart contracts using “npx hardhat compile" and deploy them locally using npx hardhat run deploy.ts. The Hardhat local node runs at 127.0.0.1:8545 and provides 20 pre-funded accounts for testing \cite{hardhat_2025}.

The voter registration is done by associating their Ethereum address with an encrypted embedding in a mapping. Initially, it will verify whether the voter is registered or not. If the voter is not registered, a new Voter struct is created with the given encrypted embeddings, and “hasVoted" is set to false, while “isRegistered" is set to true. Once this is done, it emits an event to notify that the voter has been registered successfully.
%We will use the node's address as the key and store the encrypted embedding in the struct. A struct is essentially a custom data structure that stores data related to voters. The struct is stored in the voters by mapping the key of “voterAddress".

% \begin{algorithm}
% \caption{Secure E-Voting and Blockchain Ledger Update with Falcon Signatures}
% \label{algo:voting_ledger_update}
% \begin{algorithmic}[1]

% \State  \textbf{Function} \textsc{VOTE}$(Va, Fe, CiD, Pb, Pv)$
% \State  $status \gets \Call{VERIFY}{Va, Fe, Pb}$
% \If{$status \neq \text{ValidSignature}$}
%     \State  \Return VoteRejected
% \EndIf
% \If{$voter.hasVoted = \text{true}$}
%     \State  \Return AlreadyVoted
% \EndIf
% \State  $Block \gets \Call{new}{BlockStruct}$
% \State  $Block.Voter \gets VoterAddress$
% \State  $Block.Candidate \gets CandidateID$
% \State  $Block.Timestamp \gets \Call{getTime}{}$
% \State  $BlockDigest \gets \Call{generateBlockSignature}{Block}$
% \State  $Block.Signature \gets \Call{falcon.sign}{BlockDigest, Pv}$
% \State  \Call{AppendToLedger}{Block}
% \State  $voter.hasVoted \gets \text{true}$
% \State  $candidate.vote\_count \gets candidate.vote\_count + 1$
% \State  \Return VoteCast
% \State  \textbf{End Function}

% \end{algorithmic}
% \end{algorithm}

% Algorithm~\ref{algo:voting_ledger_update} ensures 

Thus, the authenticated voters can cast their votes and record this information in the ledger, where the verify function verifies the voter’s identity whenever a voter initiates the process, which then checks whether the face embedding used during registration and voting matches or not with the help of the Public key (Pb). Once the identity and eligibility are verified, a new block is created to store the vote, which contains the voter’s address, the chosen ID, and the timestamp when the vote was casted. Once the voting is completed, the system generates a unique hash of the block and signs it using the Private key (Pv), which is then appended to the ledger, making it tamper-proof. 

Our framework enforces a strict one-vote-per-user policy by preventing duplicate submissions.
%during registration and voting through unique identity verification and vote tracking. 
Votes are recorded on a blockchain ledger, which is linked to a unique user address for traceability and accountability while preserving anonymity. Cryptographic techniques secure voter identity, keeping personal details confidential. After casting a successful vote, a vote transaction updates the vote count for the selected candidate. The total vote count for each candidate can be queried from the ledger later, enabling real-time results or vote aggregation.

\subsection{System Integration}
The frontend is a web interface that interacts with the backend via RESTful APIs. The user captures a face image, which is sent to the backend. The MobileFaceNet model generates the face embedding, which is encrypted using Falcon. The backend then communicates with the blockchain through ethers.js to call smart contract functions. Events from the blockchain are used to confirm user actions such as registration and voting. This seamless integration provides a secure and user-friendly experience where cryptographic and blockchain operations remain transparent to the end-user \cite{ethersjs_v6_2025}.

\section{Evaluation and Testing}
\label{evalutaion}
To evaluate the robustness and effectiveness of encryption and blockchain in the proposed system, various testing mechanisms were implemented. Falcon signature scheme handled the encryption layer, which was primarily tested for its strong ability to sign and verify the biometric data with a minimal latency of 250 ms for 20 concurrent requests. The generated face embeddings were digitally signed using Falcon's private key, and these signatures were subsequently tested and verified using the public key.

\subsection{ Blockchain Deployment Setup}
The blockchain component was deployed and tested in a controlled local network environment using a local blockchain setup. In this setup, each time the user casts a vote, the system treats the vote as a blockchain transaction. The transaction includes details such as the timestamp, the anonymous identity of the voter, and the selected candidate. Once the successful vote casting is completed, it is broadcasted to the local blockchain network, which is then verified by the nodes and subsequently recorded on the blockchain with a unique transaction identity.

\subsection{System Monitoring and Auditing}
Monitoring and auditing functionalities were verified using Prometheus and Grafana \cite{Ogbuefi2021Review}. Prometheus exports metrics at a periodic interval in real-time, using an exporter client installed on our centralized server, which is responsible for recording latency, encryption, decryption time, and other metrics. The exported metrics from Prometheus can be visualized graphically in Grafana by creating dynamic and custom dashboards to view and monitor the metrics as needed.

%This module allows administrators to audit the system and validate the voting process. While maintaining voter privacy, the module enables administrators to monitor system performance, verify data integrity, and track authentication logs. This auditing feature upholds transparency and accountability without compromising the confidentiality of individual votes.

\subsection{Face Recognition Module and Anti-spoofing Performance}

\begin{table}[h!]
\centering
\caption{Benchmarking MobileNetV3 Anti-spoofing Performance.}
\renewcommand{\arraystretch}{1.5} % increases row height
\setlength{\tabcolsep}{10pt}  
\begin{tabular}{|l|c|c|}
\hline
\textbf{Dataset} & \textbf{Accuracy (\%)} & \textbf{AUC Score} \\
\hline
NUAA (print) \cite{nuaa} & $98.0 \pm 0.4$ & 0.996 \\
\hline
Replay-Attack \cite{replay} & $97.0 \pm 0.7$ & 0.9896 \\
\hline
\end{tabular}
\label{auuaa}
\end{table}

AdaFace uses a quality-adaptive margin leveraging feature norm as a proxy for image quality to weight training samples, pretrained on large face datasets like MS-Celeb-1M or VGGFace2 and fine-tuned for our application. The AdaFace-based model achieved state-of-the-art verification performance in testing, outperforming prior methods on IJB-B/C and IJB-S benchmarks \cite{Kim_2022_CVPR}. This confirms that adaptive-margin learning stabilizes recognition accuracy even on low-quality images. All face embeddings are compared using cosine similarity, with a threshold tuned on a held-out set.

The classifier is trained to distinguish live faces from spoof attacks. Building on prior work, we utilized the large-scale CelebA-Spoof dataset and the MobileNetV3 model. In addition, we augmented training with samples from smaller spoofing datasets, as in Table~\ref{auuaa}, to improve generalization. The network was enhanced with a CBAM and a CDC in each block to focus on fine-grained artifact features. These architectural improvements strengthened the model’s ability to detect subtle spoofing cues. 

\section{Results and Discussion}
This section provides analytical results derived from our experiments on post-quantum Falcon encryption and anti-spoofing techniques.

\subsection{Performance of MobileNetV3-based Anti-Spoofing}
The AdaFace-based face recognition module demonstrated high verification accuracy, achieving true accept rates exceeding 98\% at low false accept rates in a verification protocol modeled after the IJB benchmark \cite{Kim_2022_CVPR}. This aligns with findings in the original AdaFace study, which reported state-of-the-art results on challenging datasets. The anti-spoofing component, built on an enhanced MobileNetV3 architecture, achieved robust performance across standard benchmarks. As shown in Table~\ref{auuaa}, the model attained over 97\% classification accuracy on both the NUAA \cite{nuaa} and Replay-Attack \cite{replay} datasets, with AUC scores of 0.996 and 0.9896, respectively. These results confirm the model’s ability to reliably distinguish between live and spoofed facial inputs, including print and replay attacks. Crucially, the model remains computationally efficient, requiring less than 0.5 GFLOPs per image, which is marginally higher than baseline MobileNetV3 while delivering improved classification performance. This accuracy and efficiency support real-time, resource-constrained deployment on edge devices in secure e-voting environments.

\begin{comment}
\begin{figure}[H]
\centering    
\includegraphics[width=\linewidth,
keepaspectratio,
trim=10pt 30pt 10pt 50pt,
clip=true]{consumption.png}
\caption{Variation of Gas consumption across multiple time frames}
\label{fig:consumption}
\end{figure}

Figure~\ref{fig:consumption} presents the percentage of gas usage across the registration 
and voting process. The gas usage for the registration process is significantly higher, 
at around 3.4\%, whereas the gas usage for the voting process remains very low, 
approximately 0.2\%. This disparity generally reflects greater computational complexity 
and data handling. Furthermore, the similar gas percentage used across three distinct 
times proposes stability and predictability under varying conditions.
\end{comment}

\subsection{Falcon Post Quantum Cryptography}
Falcon was selected by National Institute of Standards and Technology (NIST) for standardization specifically due to its compact key and signature sizes, as well as strong security \cite{falcon_sign_2025}. In practice, signature generation on a typical ARM processor (Cortex-A72 at 1.8 GHz) requires on the order of $1.0\times10^{6}$ CPU cycles for Falcon-512 and about $2.0\times10^{6}$cycles for Falcon-1024
\cite{10.1007/11761679_17}. Verification is significantly faster: Optimized tests show that Falcon’s verification throughput is roughly 3–4x higher than that of CRYSTALS-Dilithium at the same security level. %\cite{falcon_sign_2025}.
 In other words, although signing takes longer than in schemes like Dilithium (about 2x slower), 
 %\cite{10.1007/11761679_17},
 Falcon signatures can be generated and verified efficiently enough for practical use on modern hardware. Falcon scheme meets the security requirements (NIST levels 1 and 5) while maintaining high bandwidth efficiency. As noted in its design goals, the encryption results confirm the trade-offs: relatively small signature sizes at the cost of somewhat higher computation.

 % \vspace{-24pt}
% --- Table 3 ---
\begin{table}[h!]
\centering
\caption{Metrics for encryption and decryption under different loads}

\begin{tabular}{|l|c|c|}
\hline
\textbf{Metrics} & \textbf{Concurrent Request} & \textbf{Average Time} \\
\hline
\multirow{2}{*}{Encryption} & 20 & 250 ms \\
                            & 80 & 1015 ms \\
\hline
\multirow{2}{*}{Decryption} & 20 & 2.8 ms \\
                            & 80 & 10.5 ms \\
\hline
\end{tabular}%
\label{tab:encryption_decryption_metrics}
\end{table}

% --- Table 4 ---
\begin{table}[h!]
\centering
\caption{Block sizes for registration and voting transactions}
\renewcommand{\arraystretch}{1.5} % increases row height
\setlength{\tabcolsep}{12pt}      % adjust column padding

\begin{tabular}{|l|c|}
\hline
\textbf{Block Size During Transaction} & \textbf{Block Size} \\
\hline
Registration & 2275 Bytes \\
\hline
Voting       & 768 Bytes \\
\hline
\end{tabular}%
\label{tab:block_size_transaction}
\end{table}

% \begin{table}[h!]
% \centering
% \caption{Metrics for encryption and decryption and block sizes for registration and voting transactions}
% \renewcommand{\arraystretch}{1.5} % increases row height
% \setlength{\tabcolsep}{12pt}      % column padding
% \resizebox{\columnwidth}{!}{%
% \begin{tabular}{|l|c|c|}
% \hline
% \multicolumn{3}{|c|}{\textbf{Encryption / Decryption Metrics}} \\
% \hline
% \textbf{Metrics} & \textbf{Concurrent Request} & \textbf{Average Time} \\
% \hline
% \multirow{2}{*}{Encryption} & 20 & 250 ms \\
%                             & 80 & 1015 ms \\
% \hline
% \multirow{2}{*}{Decryption} & 20 & 2.8 ms \\
%                             & 80 & 10.5 ms \\
% \hline
% \multicolumn{3}{|c|}{\textbf{Block Sizes for Registration and Voting Transactions}} \\
% \hline
% \textbf{Block Size During Transaction} & \textbf{Block Size} & \\ % empty third column
% \hline
% Registration & 2275 Bytes & \\
% \hline
% Voting       & 768 Bytes & \\
% \hline
% \end{tabular}%
% }
% \label{tab:merged_table}
% \end{table}

Table~\ref{tab:encryption_decryption_metrics} presents the key performance metrics of the Falcon scheme, such as encryption and decryption times, whereas Table~\ref{tab:block_size_transaction} illustrates the block sizes and concurrency effects evaluated under various conditions to identify their efficiency. The system exhibits a non-linear escalation in processing time from 250 ms at 20 requests to 1015 ms at 80. This reflects Falcon’s sensitivity to parallel polynomial ring operations\cite{10.1007/978-3-642-30057-8_3}. During registration, block sizes reach approximately 2275 bytes, primarily due to the overhead introduced by Falcon’s signature encapsulation within the polynomial ring $\mathbb{Z}[X]/(X^{n} + 1)$. But the voting process encapsulates only essential ballot-related data, significantly reducing its block size to 768 bytes, thus achieving better throughput and minimal latency. Similarly, decryption time remains consistently low, 2.8 ms at 20 requests and 10.5 ms at 80, demonstrating Falcon’s efficiency in verification.

Falcon 512 parameter achieves NIST level 1 security by operating in the ring $\mathbb{Z}[X]/(X^{256} + 1)$ with a modulus. Falcon 1024 retains the stronger NIST Level 5 security by doubling the lattice dimension to 1024, thereby making it resistant to lattice-based attacks. The growing public size from 897 to 1793 bytes reflects the complex mathematical structure required for higher security. Computational costs also scale accordingly, with key generation time rising from 8.64 ms to 27.45 ms, and signing time increasing from 0.2-0.35 ms to 0.45-0.75 ms as the larger parameter sets at intensive levels increase polynomial multiplications along with Gaussian sampling operations \cite{10.1007/978-3-031-37679-5_18}. These metrics highlight the challenges in balancing PQC between security, key sizes, and computational efficiency. 

% \begin{figure}[H]
% \centering  
% \includegraphics[width=0.95\columnwidth, keepaspectratio, clip=true, trim=10pt 10pt 10pt 20pt,]{new_blocksize.jpeg}
% \caption{Block size for registration and voting across different time frames.}
% \label{fig:news}
% \end{figure}

\begin{figure}[H]
\centering  
\includegraphics[width=\columnwidth, keepaspectratio, clip=true]{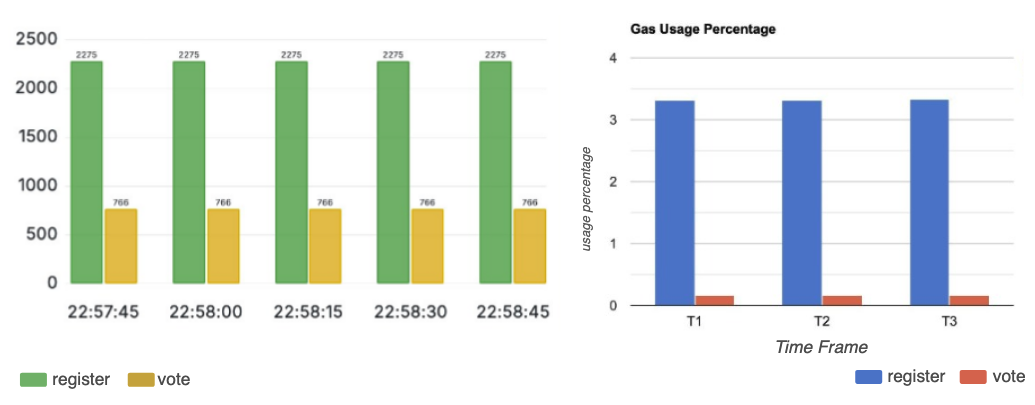}
\caption{Block size for registration and voting along with gas usage percentage.}
\label{fig:news}
\end{figure}

Figure~\ref{fig:news} compares the block size used for registration and voting. T1, T2, and T3 are the time frames during which the block size used for registration remains 2200 bytes, while the block size used for voting is significantly lower, approximately 800 bytes. The result shows registration consumes more blockchain storage than the voting process, likely due to additional data stored during onboarding.

 The percentage of gas usage across the Registration and voting process is mentioned in Figure~\ref{fig:news} . The gas usage for the registration process is significantly higher, at around 3.4\%, whereas the gas usage for the voting process remains very low, approximately 0.2\%. 

\begin{figure}[H]
    \centering
    \includegraphics[
        width=\columnwidth,
        height=!,
        keepaspectratio,
        trim=8pt 2pt 5pt 2pt,
        clip=true
    ]{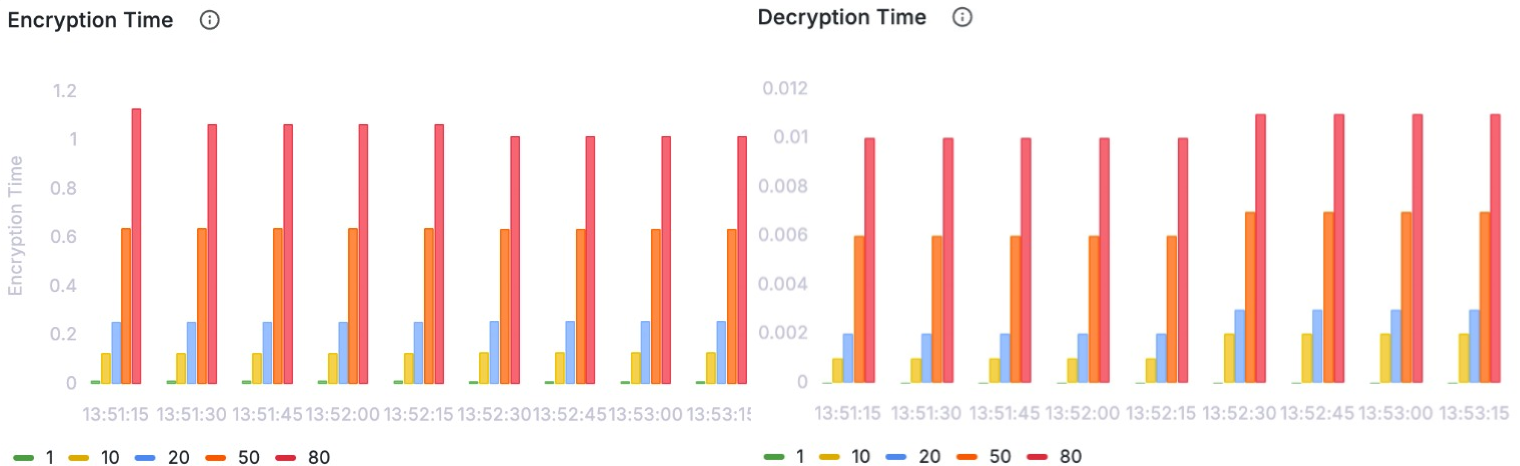}
    \caption{Impact of concurrent requests on encryption and decryption time.}
    \label{fig:impactrequest}
\end{figure}

Decryption time increases with the number of concurrent requests, indicating that the system's performance is affected during peak times. At the same time, T1 and T2 exhibit a steady increase in decryption time, with T2 performing more efficiently at lower concurrency levels. However, T3 shows irregular behavior during 10 and 20 requests. Figure~\ref{fig:impactrequest} illustrates these variations to highlight the need for potential optimization in T3 to handle more concurrent requests effectively.

\begin{comment}
\begin{figure}[H]
    \centering
    \includegraphics[width=\linewidth,
    keepaspectratio,
        trim=2pt 2pt 2pt 2pt,
        clip=true
    ]{decryption1.png}
    \caption{Variation in encryption time with increasing concurrent requests.}
    \label{fig:encryption}
\end{figure}
\end{comment}

Encryption time is demonstrated in the upper horizon, increasing with concurrent requests. T1, T2, and T3 exhibit nearly identical patterns, forecasting minimal variation. At a single request, the encryption is negligible; however, as concurrency increases, there is a sharp rise in the time taken. This uniform increase consistently prevents latency spikes and maintains stable performance during peak traffic. 

\section{Conclusion}
In this work, we develop a robust biometric authentication pipeline for e-voting. Our system integrates a MobileNetV3-based anti-spoofing module to detect presentation attacks, AdaFace for accurate facial recognition, and Falcon post-quantum encryption for enhanced security against future threats. We confirmed that this pipeline strikes a strong balance between accuracy, security, and computational efficiency. Our work demonstrates real-time biometric authentication with a latency of less than 12 ms under high concurrency, with low encryption overhead and minimal gas consumption during blockchain operation, which highlights its suitability for decentralised deployments. This validates the system's potential for secure and efficient digital democratic infrastructures.

Future research should improve biometric authentication for e-voting by ensuring lightweight, scalable, and real-time performance. Existing studies focus on isolated components, with limited blockchain and quantum encryption integration. Multi-frame analysis of short video clips can enhance liveness detection (blinks, micro-expressions) but increases computation, which lightweight temporal models (3D-CNNs, RNNs) can mitigate.

\bibliographystyle{IEEEtran}
\bibliography{references}

\end{document}